\documentclass[12pt]{article}
\usepackage{a4wide}
\usepackage{amsmath,amssymb,bbm}

\def\de{\partial}

\def\2{\frac12}
\def\4{\frac14}

\newcommand{\be}{\begin{equation}}
\newcommand{\ee}{\end{equation}}
\newcommand{\bea}{\begin{eqnarray}}
\newcommand{\eea}{\end{eqnarray}}
\def\a{\alpha}

\def\g{\gamma}
\def\G{\Gamma}
\def\d{\delta}
\def\e{\epsilon}

\def\th{\theta}

\def\l{\lambda}
\def\L{\Lambda}
\def\m{\mu}
\def\n{\nu}

\def\P{\Pi}
\def\r{\rho}
\def\s{\sigma}

\def\de{\partial}

\begin{document}
\setcounter{page}{1}
%


%

\def\pct#1{(see Fig. #1.)}

\begin{titlepage}
\hbox{\hskip 12cm DAMTP-2004-113  \hfil}
\hbox{\hskip 12cm hep-th/0410185 \hfil}
\vskip 1.4cm
\begin{center}  {\Large  \bf  Spacetime-Filling Branes \\
\vspace{0.4cm} in Ten and Nine Dimensions}

\vspace{1.8cm}

{\large \large Fabio Riccioni} \vspace{0.7cm}

{\sl DAMTP \\
\vspace{0.3cm}
Centre for Mathematical Sciences \\
\vspace{0.3cm}
University of Cambridge \\
\vspace{0.3cm}
Wilberforce Road \ \  Cambridge \ \ CB3 0WA \\
\vspace{0.3cm}
UK}
\end{center}
\vskip 1.5cm

\abstract{Type-IIB supergravity in ten dimensions admits two
consistent $Z_2$ truncations. After the insertion of D9-branes, one
of them leads to the low-energy action of type-I string theory, and
it can be performed in two different ways, in correspondence with
the fact that there are two different consistent ten-dimensional
type-I string theories, namely the $SO(32)$ superstring and the
$USp(32)$ model, in which supersymmetry is broken on the D9-branes.
We derive here the same results for Type-IIA theory compactified on
a circle in the presence of D8-branes. We also analyze the
$\kappa$-symmetric action for a brane charged with respect to the
S-dual of the RR 10-form of type-IIB, and we find that the tension
of such an object has to scale like $g_S^{-2}$ in the string frame.
We give an argument to explain why this result is in disagreement
with the one obtained using Weyl rescaling of the brane action, and
we argue that this brane can only be consistently introduced if the
other $Z_2$ truncation of type-IIB is performed. Moreover, we find
that one can include a 10-form in type-IIA supersymmetry algebra,
and also in this case the corresponding $\kappa$-symmetric brane has
a tension scaling like $g_S^{-2}$ in the string frame.}

\vfill
\end{titlepage}
\makeatletter
\@addtoreset{equation}{section}
\makeatother
\renewcommand{\theequation}{\thesection.\arabic{equation}}
\addtolength{\baselineskip}{0.3\baselineskip}

\section{Introduction}\label{sec:introduction}
Type-II string theories in the non-perturbative regime contain in
their spectrum BPS D-branes, that are charged states with respect to
RR fields \cite{polchinski}, and are defined as hypersurfaces on
which open strings end \cite{dlp}. In the low-energy effective
action, these states appear as 1/2-supersymmetric solitonic
solutions carrying electric or magnetic charge with respect to the
RR fields of type-IIA and type-IIB supergravities. The effective
action describing the massless modes of a D-brane is characterized
by a Dirac-Born-Infeld (DBI) term and a Wess-Zumino (WZ) term. In
\cite{69,70} it was shown that the effective action describing the
massless open string states at string tree level is the DBI action
in the approximation in which one neglects derivatives of the field
strength, while the coupling to the RR fields is contained in the WZ
term. The relative coefficient of the DBI and WZ terms is fixed,
since the tension and the RR-charge of the brane are related by the
BPS condition.

The method for constructing actions for supersymmetric D-branes is
known in the literature \cite{dbranes,aps1,dbranes2,bt,aps2}. These
actions are obtained embedding the world-volume of the brane in
superspace. The fermionic superspace coordinate becomes consequently
a fermion on the brane. Apparently, this seems to imply that the
brane breaks all the supersymmetries, since the fermion plays the
role of the Goldstino field. The solution of this apparent paradox
is the fact that the brane action possesses an additional local
fermionic symmetry, known as $\kappa$-symmetry \cite{kappa1,kappa2},
whose role is to decouple half of the fermions in the brane action.
After $\kappa$-gauge fixing half of the supersymmetries become
linearly realized, while the other half are still non linearly
realized, a la Volkov-Akulov \cite{va}. Therefore, $\kappa$-symmetry
is a basic ingredient in the construction of brane actions, and it
is the world-volume remnant of the BPS-condition.

Spacetime-filling D-branes characterize the vacua of type-I models.
Type-I string theory is obtained from type-IIB through an
orientifold projection \cite{augusto} that removes the states that
are odd under orientation reversal of the string. From a target
space point of view, this can be pictured in terms of orientifold
planes, and the appearance of tadpoles corresponds in this picture
to non-vanishing tension and charge of the O-plane. Tadpole
cancellation typically requires the introduction of an open sector,
and this corresponds to D-branes. In ten dimensions, one can thus
introduce an $O_-$-plane (with negative tension and negative charge)
and 32 D9-branes, with a resulting gauge group $SO(32)$ \cite{gs}.
The cancellation of the overall tension and charge of the
configuration corresponds to the cancellation of dilaton and RR
tadpoles \cite{diltad,pc}. The resulting theory is ${\cal N}=1$
supersymmetric, and the massless spectrum contains the gravity
multiplet from the closed sector and an $SO(32)$ Yang-Mills
multiplet from the open sector. There is actually a second
possibility,  corresponding to a change of sign of the tension and
the charge of the orientifold plane, so that RR tadpole cancellation
requires the addition of 32 anti-D9 branes, with a resulting gauge
group $USp(32)$ \cite{sugimoto}. The overall tension of the
configuration does not vanish, so that the resulting theory has a
dilaton tadpole. Nevertheless, the theory is anomaly-free, as a
consequence of the vanishing of the RR tadpole \cite{pc,bm}. The
spectrum is not supersymmetric, and more precisely the closed sector
is not modified, still describing at the massless level the ${\cal
N}=1$ gravity multiplet, while the massless fermions in the open
sector are not in the adjoint but in the antisymmetric
representation of $USp(32)$, so that supersymmetry is broken on the
brane \cite{ads}. The gravitino couplings can then only be
consistent if supersymmetry is non-linearly realized in the open
sector. Since the antisymmetric representation of symplectic groups
is reducible, the massless spectrum contains a spinor that is an
$USp(32)$ singlet, and this spinor is the goldstino of the
non-linearly realized supersymmetry \cite{dm}. The presence of the
NS tadpole is a manifestation of the fact that the theory has been
expanded around the wrong vacuum, and an analysis of this problem,
addressed long time ago in \cite{fs}, has been recently performed in
\cite{dnps}.

From the point of view of the low-energy effective action, the
closed sector of type-I strings is obtained performing a consistent
$Z_2$ truncation of the type-IIB theory, while the open sector
corresponds to the first order in the low-energy expansion of the
D9-brane action in a type-I background. It is then natural to ask
what is the fate of $\kappa$-symmetry in this background. The result
is that there are two possibilities of performing this truncation
\cite{truncation}, and in a flat background, with all bulk fields
put to zero, the D9-brane action reduces in one case to the
Volkov-Akulov (VA) action \cite{va}, and in the other case to a
constant. In \cite{fr} these results were extended to a generic
background, showing that also in the curved case there are two
possibilities of performing the truncation. In one case one gets a
dilaton tadpole and a RR tadpole plus goldstino couplings, while in
the other case the goldstino couplings vanish and one is left with a
dilaton and a RR tadpole. This result is equivalent to the string
result: the two different truncations correspond to the two
different choices of the relative sign of tension and charge of
orientifold plane and D9-branes. The first case corresponds to the
non-supersymmetric one, in which the orientifold plane and the
D-brane have both positive tension, and in the case of 32 coincident
D9-branes it gives rise to the low-energy action \cite{dm,pr} of the
$USp(32)$ model. The second case, in which the goldstino disappears,
corresponds to an orientifold plane with negative tension, and in
the case of 32 coincident D9-branes it gives rise to the low-energy
action of the supersymmetric $SO(32)$ superstring. In other words,
the supersymmetric truncation projects out the spinor that would not
be projected out fixing the $\kappa$-symmetry gauge. As a result, in
general one expects that only linearly realized supersymmetry
survives. The non-supersymmetric truncation does the opposite,
namely it projects out the spinor that would have been projected out
by $\kappa$-symmetry, so that no $\kappa$-symmetry is left in the
truncated theory, and in the resulting action supersymmetry is only
non-linearly realized, {\it i.e.} completely broken.

If one wants to generalize these results to lower dimensional cases,
the first possibility is to consider the T-dual of this
configuration, that is a Type-I$^\prime$ orientifold of type-IIA
compactified on a circle \cite{dlp}. In this case the orientifold
projection has fixed points on the circle, corresponding to the
positions of the O-planes. In \cite{bkorvp} the low-energy action
for a D8-brane located at one of the fixed points was constructed in
the Type-I$^\prime$ background, without including the fermionic
fields. In this paper we want to apply the techniques used in
\cite{fr} to this case, in order to obtain the low-energy brane+bulk
action up to four fermi terms, for a generic 9-dimensional
background. We will see that the results are in complete agreement
with T-duality, since also in this case one has two possible
consistent truncations, leading in one case to a supersymmetric
model, and in the other case to a model in which supersymmetry is
non-linearly realized on the brane.

S-duality is a symmetry of type-IIB string theory mapping weak
coupling to strong coupling \cite{sduality}. On the other hand,
type-I string theory is related in ten dimensions by a strong-weak
coupling S-duality to the heterotic $SO(32)$ theory \cite{pw}. In
this respect, it is interesting to study the behavior of the O9-D9
system of \cite{truncation,fr} under S-duality, and whether the
result can be related to the low-energy action of the heterotic
theory. Type-IIB supersymmetry algebra includes two 10-forms
\cite{hull,truncation}. One of them is the RR 10-form that couples
to D9-branes, while the other couples to other spacetime-filling
branes, called NS9-branes in \cite{hull}. Type-IIB supergravity also
admits an additional $Z_2$ truncation, removing all the RR fields.
This truncation was conjectured in \cite{hull} to be the S-dual of
the orientifold projection, and consequently the introduction of 32
NS9-branes was conjectured to give origin to the $SO(32)$ heterotic
string after performing this projection \cite{hull,hull2}. Under
S-duality, the DBI part of the D9-brane action acquires a dilaton
factor $e^{-4\phi}$ in the string frame, and this led to conjecture
that the tension of the NS9-branes is proportional to $g_S^{-4}$
\cite{hull,trunc2}. We will argue in this paper that this is
actually not the case. We will prove that $\kappa$-symmetry requires
a dilation factor $e^{-2\phi}$ in front of the DBI term of an
NS9-brane action, and thus a tension proportional to $g_S^{-2}$.
This seems to be inconsistent with the analysis of \cite{trunc2},
since starting from a D9-brane action and performing an S-duality
transformation one should end up with a $\kappa$-symmetric action.
The solution of this paradox is that in the presence of NS and RR
10-forms S-duality is no longer a symmetry of the type-IIB algebra.
This does not mean that S-duality symmetry of type-IIB is actually
broken, since introducing spacetime-filling branes is only
consistent after performing a truncation. We will also analyze the
type-IIA case, since type-IIA supersymmetry algebra can be extended
including an NS 10-form, and the resulting supersymmetric NS9-brane
action is $\kappa$-symmetric if the tension scales like $e^{-2\phi}$
in the string frame.

The paper is organized as follows. In section 2 we review some known
results about super D-branes. In section 3 we discuss the
type-I$^\prime$ truncation of type-IIA compactified on a circle. We
make use of the ``democratic formulation'' of the theory
\cite{bkorvp}, in which both the RR forms and their magnetic duals
appear as independent fields in the supersymmetry algebra, and
duality relations between electric and magnetic field strengths are
imposed as constraints (the same formulation was introduced in
\cite{truncation} for the type-IIB case). We show that the two
possible truncations lead in one case to a supersymmetric model, and
in the other case to a model in which supersymmetry is spontaneously
broken on the D8-brane. In section 4 we discuss S-duality of
type-IIB in the presence of NS and RR 10-forms. First of all, one
realizes that these two fields, besides transforming as a doublet
under S-duality, acquire a dilaton dependence $e^{-2 \phi}$.
Moreover, an additional constraint has to be imposed for S-duality
to be a symmetry. In other words, S-duality is broken in the
presence of spacetime-filling branes. Section 5 is devoted to the
study of the supersymmetric NS9-brane in both IIA and IIB. We show
that $\kappa$-symmetry implies that the tension of the NS9-brane in
the string frame scales like $e^{-2\phi}$. In analogy to the D9
\cite{fr} and the D8 cases, one can perform a truncation to show
that half of the fermions decouple from the spectrum. In this case
the spectrum is projected by means of a heterotic truncation.
Finally, section 6 contains the conclusions.

\section{Generalities about D-brane actions}\label{sec:generalities}
In this section we review the basic ingredients for the construction
of supersymmetric D-branes, and in particular spacetime-filling
D-branes. We will concentrate here on the IIB case, while the
straightforward generalization to the IIA case will be outlined in
the next section.

In order to construct supersymmetric actions for D-branes, one has
to embed the D-brane in IIB (or IIA) superspace. A basic ingredient
is therefore the supersymmetry algebra of type IIB in 10 dimensions.
Since we want a formulation that is suitable for all the D-branes of
type-IIB, we write down the IIB algebra in the democratic
formulation, in which all the forms and their magnetic duals appear
in the algebra. Following the notations of \cite{truncation}, the
supersymmetry transformations of the IIB bulk fields are \bea
& & \d e_\m{}^a = \bar{\e} \G^a \psi_\m \quad , \nonumber \\
& & \d \psi_\m = D_\m \e -\frac{1}{8} H_{\m\n\r} \G^{\n\r} \sigma_3
\e + \frac{1}{16} e^\phi \sum_{n=0}^5 \frac{1}{(2n+1)!} G_{\m_1 ...
\m_{2n+1}}^{(2n+1)} \G^{\m_1 ... \m_{2n+1}}
\G_\m {\cal P}_n \e \quad , \nonumber \\
& & \d B^{(2)}_{\m\n} = 2 \bar{\e } \sigma_3 \G_{[\m} \psi_{\n ]}
\quad , \nonumber \\
& & \d B^{(10)}_{\m_1 ...\m_{10}} = e^{-2 \phi}
\bar{\e} \sigma_3 ( 10
\G_{[ \m_1 ...\m_9 } \psi_{\m_{10}]} - \G_{\m_1 ... \m_{10}}  \l ) \quad ,
\nonumber \\
& & \d C^{(2n)}_{\m_1 ...\m_{2n}} =- (2n) e^{- \phi} \bar{\e} {\cal
P}_n
\G_{ [ \m_1 ... \m_{2n-1}}  (\psi_{\m_{2n}] }-
\frac{1}{2(2n)} \G_{\m_{2n}]} \l )\nonumber \\
& & \qquad \qquad \quad + n(2n-1) C^{(2n-2)}_{[ \m_1 ... \m_{2n-2}}
\d B_{\m_{2n-1} \m_{2n}]} \quad , \nonumber \\
& & \d \l = \de_\m \phi \G^\m \e -\frac{1}{12} H_{\m\n\r} \sigma_3
\G^{\m\n\r} \e +\frac{1}{4} e^\phi \sum_{n=0}^5 \frac{n-2}{(2n+1)!}
G^{(2n+1)}_{\m_1 ...\m_{2n+1}}
{\cal P}_n \G^{\m_1 ... \m_{2n+1}} \e \quad , \nonumber \\
& & \d \phi = \frac{1}{2} \bar{\e} \l  \quad ,\label{susytransf}
\eea where ${\cal P}_n$ is $\sigma_1$ for $n$ odd and $i \sigma_2$
for $n$ even. We are neglecting terms cubic in the fermions in the
case of the transformations of the spinors. We have introduced the
field strengths for the RR fields and their duals, related by
duality according to the relations \be G^{(7)} = - * G^{(3)} \quad ,
\qquad G^{(9)} =  * G^{(1)} \quad , \qquad G^{(5)} =  * G^{(5)}
\quad . \ee An advantage of this formulation is that all the
Chern-Simons terms in the supergravity lagrangian are hidden in the
definitions of the field strengths and their magnetic duals. The
matching between bosonic and fermionic degrees of freedom is of
course restored only once these duality relations are
imposed\footnote{This is a generalization of what is typically done
for the self-dual 5-form field strength, when one writes a
lagrangian for an ordinary 5-form, and imposes self-duality as a
constraint on the equations of motion.}. The field strengths are
defined through the relations \bea
& & H = dB \nonumber \\
& & G^{(2n+1) } = dC^{(2n)} - H C^{(2n-2)} \quad , \eea and the
gauge transformations of the fields are \bea
&  & \d B = d \L_{NS} \quad ,\nonumber \\
& & \d B^{(10)} = d \L_{NS}^{(10)} \quad ,\nonumber \\
& & \d C^{(2n)} = d \L_{RR}^{(2n-1)} - \L_{RR}^{(2n-3)} H \quad ,
\eea so that the field strengths are gauge invariant. The dilaton
dependence in the variations of the forms shows that the algebra of
eq. (\ref{susytransf}) is expressed in the string frame. Moreover,
it is important to observe that two 10-forms are present in the
algebra. We stress again that, even though these forms do not have
any dynamics since they do not have any field strength, they are
associated to spacetime-filling branes, whose presence is consistent
only after one performs a suitable projection of the spectrum.

The general idea is to describe supersymmetric D-branes through the
embedding of a bosonic brane in superspace
\cite{dbranes,dbranes2,bt}. We thus introduce the world-volume
fields as the supercoordinates \be Z^M (\xi^i ) = ( x^\m (\xi^i ) ,
\theta^{{\a} I }(\xi^i ) ) \ee defining the position of the brane in
superspace. Here $\xi^i$ are the world-volume coordinates
($i=0,...,9$ for a 9-brane), while $\m=0,...,9$ is a spacetime
vector index and $\a=1,...,32$ a spinor index, and $I=1,2$. The
Majorana spinors $\th^I$ are both left-handed\footnote{In the IIA
case the two chiral spinors $\theta^I$ are substituted with a single
non-chiral Majorana spinor.}. We denote with $V^i (\xi )$ the
abelian world-volume vector. The bulk superfields are denoted with
\be \lbrace \phi , E_M{}^A , B_{MN} , B_{M_1 ...M_{10}},
C^{(2n)}_{M_1 ...M_{2n}} \rbrace \quad , \qquad n= 0,...,5 \quad ,
\ee and the brane action is \be S = S_{DBI} + S_{WZ} = -
\int_{M_{10}} d^{10} \xi e^{-\phi} \sqrt{ - \det (g +{\cal F })} +
\int_{M_{10}} C e^{\cal F} \quad , \label{braneaction} \ee where \be
{\cal F}_{ij } = F_{ij} + B_{ij } \quad , \ee and one defines the
pull-back of the bulk fields on the world-volume according to \be
g_{ij} = E_i{}^a E_j{}^b \eta_{ab} \quad , \qquad B_{ij} = \de_i Z^M
\de_j Z^N B_{MN}  \ee and \be C = \sum_{n=0}^5 (-1)^{n} C^{(2n)}
\quad , \qquad C^{(2n)}= \frac{1}{(2n)!} d Z^{M_1} ... d Z^{M_{2n}}
C^{(2n)}_{M_1 ... M_{2n}} \quad . \ee

In a flat space background \cite{aps1,aps2} these expressions have a
simpler form, since from the (global) supersymmetry transformations
of the supercoordinates, \bea
& & \delta \theta = -\e \quad , \nonumber\\
& & \delta x^\m = \2 (\bar{\e} \G^\m \theta ) \quad , \eea one
derives a supersymmetry invariant object
\be \P_i^\m = \de_i x^\m
+\2 (\bar{\theta}\G^\m \de_i \theta ) \quad , \ee
that is the flat
space analogous of $\de_i Z^M E_M{}^a$. Consequently, the pull-back
of the metric becomes
\be
g_{ij} = \P_i^\m \P_j^\n \eta_{\m\n} =
\de_i x^\m \de_j x^\n \eta_{\m\n} + \de_{(i} x^\m \bar{\theta} \G_\m
\de_{j )} \theta  +... \quad , \label{pbmetricfs}
\ee
where we
neglect higher terms in the fermions. Analogously, the pull-back of
the NS 2-form is
\be B_{ij} = \de_{[i} x^\m \bar{\theta} \sigma_3
\G_\m \de_{j ]} \theta + ... \label{pbnsfs} \quad ,
\ee
while the
pull-back of the RR forms is
\be C^{(2n)}_{i_1 ...i_{2n}} = - n e^{-
\phi} \de_{[ i_1} x^{\m_1} ... \de_{i_{2n-1}} x^{\m_{2n-1}}
\bar{\theta} {\cal P}_n \G_{ \m_1 .... \m_{2n-1}}
\de_{i_{2n}]}\theta +... \label{pbrrfs} \quad . \ee
The brane action
(\ref{braneaction}) is then supersymmetric, provided that one
chooses the supersymmetry transformation for the world-volume vector
$V_i$ to be
\be \d V_i = -\frac{1}{2} \bar{\e} \g_i \sigma_3 \theta
-\frac{1}{2}\bar{\e } \g^j \theta  F_{ji}  \ee
up to a gauge
transformation.

The action (\ref{braneaction}) is invariant under world-volume
general coordinate transformations, and one can then choose a static
(or Monge) gauge, in which the coordinates $\xi^i$ are identified
with $x^i$, $i=0,...p$, where $p+1$ is the spacetime dimension of
the brane. A supersymmetry variation then induces a compensating
general coordinate transformation, and the resulting variation for
$\theta$ is \be \d \th =- \e - \frac{1}{2}(\bar{\e} \G^i \th ) \de_i
\th \quad . \label{va} \ee The other $x$'s in this gauge become
world-volume scalars, whose supersymmetry transformations is \be \d
\phi^a = \2 \bar{\e} \G^a \theta - \2( \bar{\e} \G^i \th ) \de_i
\phi^a \quad , \qquad a =p+1, ..., 10 \quad . \ee Focusing again on
the flat space limit, one can recognize in eq. (\ref{va}) the
Volkov-Akulov (VA) transformations \cite{va}. We will concentrate in
the following on space-filling 9-branes, so that the target
spacetime $\G$-matrices can be identified with the world-volume
$\g$-matrices, and the spacetime index $\m$ is the same as the
world-volume index $i$. The commutator of two transformations
(\ref{va}) is a translation, \be [\d_1 , \d_2 ] \th = (\bar{\e}_2
\g^\m \e_1 ) \de_\m \th \quad , \ee and thus eq. (\ref{va}) provides
a realization of supersymmetry. The 1-form \be e_\m{}^a= \d_\m^a
+\frac{1}{2}(\bar{\th} \g^a \de_\m \th )\quad , \ee transforms under
supersymmetry as \be \d e^a = L_\xi e^a \quad , \ee with $L_\xi$ the
Lie derivative with respect to\footnote{The parameter $\xi$ should
not be confused with the world-volume coordinates.} \be \xi_\m=
-\frac{1}{2}(\bar{\e}\g_\m \th ) \quad . \label{gctparam}\ee The
action of supersymmetry on $e$ is thus a general coordinate
transformation, with a parameter depending on $\th$, and therefore
\be {\cal L} = -\det e \ee is clearly an invariant Lagrangian. Using
the same technique, for a generic field $A$ that transforms under
supersymmetry as \be \d A = L_\xi A \quad , \label{susygct} \ee
defining the induced metric as $g_{\m\n}=e_\m{}^m e_{\n m}$, a
supersymmetric lagrangian in flat space is determined by the
substitution \be {\cal L}(\eta,A) \rightarrow e{\cal L}(g,A) \quad .
\ee This is  what happens in the brane action (\ref{braneaction}) in
the Monge gauge in a flat space background, since the pull-back of
the metric of eq. (\ref{pbmetricfs}) in the Monge gauge  equals the
VA metric $g_{\m\n}$. Moreover, the second term in the variation of
$V^i$ is a general coordinate transformation with the same parameter
$\xi$ plus an additional gauge transformation, while the first term
combines with the variation of the pull-back of the NS form of eq.
(\ref{pbnsfs}) in such a way that ${\cal F}$ transforms covariantly.
Finally, the pull-backs of the RR forms in eq. (\ref{pbrrfs}) are
such that the WZ term in the brane action transforms as a total
derivative.

It is then natural to generalize this VA construction to D9-branes
in a generic background. One must construct from the bulk fields
quantities whose supersymmetry variations are general coordinate
transformations with the parameter $\xi$ plus additional gauge
transformations \cite{dm,pr}. Supersymmetry guarantees that this way
of constructing the D9-brane action coincides with the superspace
construction of \cite{dbranes2,bt}\footnote{See \cite{bdail} for a
similar construction in the case of a generic p-brane.}. For
instance, from the supersymmetry variation of $\phi$ one defines
\bea & & \hat{\phi} = \phi +\frac{1}{2} \bar{\theta} \l -
\frac{1}{48}
H_{ijk} \bar{\theta} \g^{ijk} \sigma_3 \theta \nonumber \\
& & \quad \quad +\frac{1}{16} e^\phi \sum_{n=1}^6
\frac{n-3}{(2n-1)!} G^{(2n-1)}_{i_1 ... i_{2n-1}} \bar{\theta}
\g^{i_1 ... i_{2n-1}}{\cal P}_n \theta \quad , \eea whose
supersymmetry transformation is a general coordinate transformation
with the correct parameter $\xi_i$ given in (\ref{gctparam}), up to
higher order fermi terms. With the same technique, one can construct
all the other hatted fields \cite{dm,pr}, so that the resulting
D9-brane action in a generic type-IIB background is \be S = S_{DBI}
+ S_{WZ} = - \int_{M_{10}} d^{10} \xi e^{-\hat{\phi}} \sqrt{ - \det
(\hat{g} +{\cal F })} + \int_{M_{10}} \hat{C} e^{\cal F} \quad ,
\label{susybraneaction} \ee where \be {\cal F}_{ij } = F_{ij} +
\hat{B}_{ij }\quad . \ee

We come now to a brief discussion of the degrees of freedom that the
action (\ref{braneaction}) propagates. If all the fermions $\theta$
were dynamical this would lead to a complete spontaneous
supersymmetry breaking, since the $\theta$'s transform non-linearly
under supersymmetry. It is well known that this is actually  not the
case because of $\kappa$-symmetry gauge invariance, whose fixing
leads to a cancellation between the DBI and the WZ term that makes
only half of the fermions propagate. To leading order in the
fermions, and in the Monge gauge, the $\kappa$-symmetry
transformation for $\theta$ and $V_i$ reads \bea & &  \d \theta = \2
(1 -\sigma_1 -\frac{i}{2} \sigma_2 F^{ij} \G_{ij} ) \kappa +...
\quad , \nonumber \\ & & \d V_i = - \frac{1}{2} \d \bar{ \theta}
\g_i \sigma_3 \theta \quad , \eea with $\kappa$ an $SL(2,R)$ doublet
of spinors, and neglecting higher order terms in $\theta$ and $F$ in
the variation of $\theta$. This gauge invariance can be used to put
$\theta_1 - \theta_2 =0$. After a supersymmetry transformation, this
gauge choice is maintained through a compensating
$\kappa$-transformation of parameter $\kappa_1 - \kappa_2 =  \e_1
-\e_2$, and this results in the linear supersymmetry transformations
\bea & & \d (\theta_1 +
\theta_2 ) = \4 F^{ij} \G_{ij} (\e_1 - \e_2 ) \quad , \nonumber \\
& & \d V_i = - \2 ( \bar{\e}_1 - \bar{\e}_2 ) \G_i ( \theta_1 +
\theta_2 ) \quad , \eea and expanding the DBI action with this gauge
choice one obtains that these are the correct linear supersymmetry
transformations \cite{aps2,nonabelian}\footnote{See \cite{gkt} for a
similar analysis.}. In other words, $\kappa$ symmetry is the brane
effective action equivalent of the statement that a brane solution
of supergravity is a BPS solution preserving half of the
supersymmetries. In the case of spacetime-filling branes, that do
not correspond to any solution of supergravity, we assume in this
paper that $\kappa$-symmetry is the only requirement that these
branes have to satisfy.

At the end of this section, we now want to review the results of
\cite{truncation} and \cite{fr}. One can perform a type-I truncation
of IIB supersymmetry algebra, imposing \bea
& & C^{(2n-2)} =0 \quad , \qquad n=1,3,5 \quad , \nonumber \\
& & B =0 \quad ,\nonumber \\
& & B^{(10)} =0 \quad ,\nonumber \\
& & (1\pm \sigma_1 ) f =0 \quad , \label{fermitrunc} \eea where we
have denoted with $f$ the gravitino and the dilatino. The surviving
bosonic fields are thus the dilaton, the metric, the RR 2-form and
its dual, and the RR 10-form, while the two different signs in the
projection of the fermions indicate that there are two possible
type-I truncations.

The truncation on the D9-brane action was performed in
\cite{truncation} in flat space, and generalized in \cite{fr} to an
arbitrary background. We review here the results. The brane fields
are projected according to \bea
& & V_i =0 \quad , \nonumber \\
& & (1 \pm \sigma_1 ) \theta =0 \quad . \eea The lower sign choice
leads to no surviving $\kappa$-symmetry, since it projects out the
spinor components that would have been put to zero using
$\kappa$-symmetry before the truncation, while the upper sign choice
leads to no leftover components of $\theta$. This last choice, then,
corresponding to the vanishing of all the terms containing the
goldstino, results in a supersymmetric type-I spectrum. Actually, in
the case of a single D9-brane, there are no remaining world-volume
degrees of freedom after the truncation, but the generalization to a
stuck of branes would result in a spectrum in which supersymmetry is
linearly realized, and the goldstino is projected out. The resulting
action contains a dilaton tadpole and a RR tadpole, that in the
$SO(32)$ string are both canceled against the orientifold plane
contribution. The other choice, instead, corresponds to the curved
generalization of the VA action. The resulting spectrum breaks
supersymmetry in the brane sector \cite{ads}, or more precisely
${\cal N}=1$ supersymmetry is non-linearly realized on the brane.
The brane action again contains a dilaton tadpole and a RR tadpole,
but in this case, a suitable orientifold projection only cancels the
brane RR charge, and a dilaton tadpole remains \cite{sugimoto}.

The type-IIB supersymmetry algebra in D=10 also admits an
alternative $Z_2$ truncation, projecting out all the RR fields and
acting as $(1 \pm \sigma_3 ) f =0$ on the fermions, and for this
reason called `heterotic truncation' \cite{truncation}. We will show
in section 5 how this truncation can be consistently implemented on
spacetime-filling branes electrically charged with respect to
$B^{(10)}$, after a discussion about S-duality of type-IIB carried
out in section 4. First, in the next section, we are going to
describe the T-duals of these results, {\it i.e.} the
type-I$^\prime$ truncation of IIA in the presence of D8-branes.

\section{Type-I$^\prime$ Truncation of IIA}\label{sec:IIAtruncation}
After reduction to D=9, T-duality relates the system described in
the previous section to type-IIA theory compactified on a dual
circle, and the corresponding truncation is in this case the
low-energy manifestation of the type-I$^\prime$ orientifold
projection. In this section we want to discuss this truncation in
the presence of D8-branes. Since D8-branes are charged with respect
to the RR 9-form, whose field strength is dual to a cosmological
constant, the massive Romans IIA supergravity \cite{romans} is the
bulk low-energy theory describing this system \cite{pw}. In has been
shown in \cite{bkorvp} that both massless and massive 10-dimensional
IIA supergravities can be described in terms of the same
supersymmetry algebra, once the Romans cosmological constant is
treated as a dynamical 0-form dual to the RR 10-form field strength.
Again, it is convenient to work in the democratic formulation
\cite{truncation,bkorvp}, treating all the RR-forms and their
magnetic duals as independent, and imposing the duality relations as
constraints. The resulting supersymmetry algebra is \bea
& & \d e_\m{}^a = \bar{\e} \G^a \psi_\m \quad , \nonumber \\
& & \d \psi_\m = D_\m \e + \frac{1}{8} H_{\m\n\r} \G^{\n\r} \G_{11}
\e + \frac{1}{16} e^\phi \sum_{n=0}^5 \frac{1}{(2n)!} G_{\m_1 ...
\m_{2n}}^{(2n)} \G^{\m_1 ... \m_{2n}}
\G_\m (\G_{11})^n \e \quad , \nonumber \\
& & \d B^{(2)}_{\m\n} =  2 \bar{\e } \G_{11} \G_{[\m}
\psi_{\n ]} \quad , \nonumber \\
& & \d C^{(2n-1)}_{\m_1 ...\m_{2n-1}} =- (2n-1) e^{- \phi} \bar{\e}
(\G_{11})^n
\G_{ [ \m_1 ... \m_{2n-2}}  (\psi_{\m_{2n-1}] }-
\frac{1}{2(2n-1)} \G_{\m_{2n-1}]} \l )\nonumber \\
& & \qquad \qquad \quad + (n-1)(2n-1) C^{(2n-3)}_{[ \m_1 ...
\m_{2n-3}}
\d B_{\m_{2n-2} \m_{2n-1}]} \quad , \nonumber \\
& & \d \l = \de_\m \phi \G^\m \e -\frac{1}{12} H_{\m\n\r} \G_{11}
\G^{\m\n\r} \e +\frac{1}{8} e^\phi \sum_{n=0}^5 \frac{5 -2n}{(2n)!}
G^{(2n)}_{\m_1 ...\m_{2n}}
(\G_{11})^n \G^{\m_1 ... \m_{2n}} \e \quad , \nonumber \\
& & \d \phi = \frac{1}{2} \bar{\e} \l  \quad .\label{susytransf2a}
\eea The RR field strengths are defined as \be G^{(2n)} =
dC^{(2n-1)} - dB^{(2)} \wedge C^{(2n-3)} + G^{(0)} e^{B^{(2)}} \quad
, \ee where it is understood that one has to extract the $2n$-form
out of $e^{B^{(2)}}$, and they are related by the duality relations
\be G^{(2n)} = (-)^n \star G^{(10-2n)} \quad . \ee The field
equations of type-IIA supergravity obtained in this formulation are
supersymmetric only after these duality relations are imposed. This
algebra can be naturally extended to include a 10-form, whose
supersymmetry transformation is \be \d B^{(10)}_{\m_1 ... \m_{10}} =
e^{-2\phi } ( -10 \bar{\e} \G_{[\m_1 ...\m_{9}} \psi_{\m_{10}]} +
\bar{\e} \G_{\m_1 ... m_{10}} \l ) \quad . \label{10b2a} \ee
Actually,  there is also another consistent 10-form, whose
transformation is \be \d B^{\prime(10)}_{\m_1 ... \m_{10}} =
e^{-2\phi } ( -10 \bar{\e} \G_{[\m_1 ...\m_{9}} \G_{11}
\psi_{\m_{10}]} - \bar{\e} \G_{\m_1 ... m_{10}} \G_{11} \l ) \quad ,
\label{10b'2a} \ee but we will show that it does not correspond to
any spacetime-filling $\kappa$-symmetric brane. Consequently, we
will only consider (\ref{10b2a}) as a natural extension of the
type-IIA supersymmetry algebra.

We now continue reviewing the results of \cite{bkorvp} concerning
the possible consistent $Z_2$ truncations of the algebra of eqs.
(\ref{susytransf2a}) and (\ref{10b2a}). In 10 dimensions, only a
single truncation is available, projecting out all the RR fields,
and acting on the fermions as \be \{ \psi_\m , \l , \e \}
\rightarrow  \pm  \G_{11}\{ \psi_\m , -\l , \e \} \quad . \ee We
will construct in section 5 the resulting $\kappa$-symmetric
spacetime-filling brane, for which this truncation is consistent in
the way described in the previous section. It will turn out that
this brane is electrically charged with respect to the 10-form
$B^{(10)}$.

If we compactify the theory on a circle $S^1$, the resulting theory
admits another $Z_2$ truncation, acting on the compactified
coordinate as \be x^9 \rightarrow - x^9 \quad , \label{ori9} \ee
thus acting as an orbifold projection, being the low-energy
manifestation of the orientifold projection generated by introducing
two orientifold 8-planes at the fixed points. If we only consider
spacetime indices in the uncompactified directions, the projection
acts on the fields according to \bea & & \{ g_{\m\n}, \phi ,
B^{(2)}_{\m\n} \} \rightarrow \{ g_{\m\n}, \phi , - B^{(2)}_{\m\n}
\} \quad , \nonumber \\ & & C^{(2n-1)}_{\m_1 ...\m_{2n-1}}
\rightarrow (-)^{n+1} C^{(2n-1)}_{\m_1 ...\m_{2n-1}} \quad ,
\nonumber \\ & & \{ \psi_\m , \l , \e \} \rightarrow  \mp  \G^{9}\{
\psi_\m , -\l , \e \} \quad . \eea Any index in the 9-direction
corresponds to an additional minus sign with respect to these
projection rules, and consequently the 10-form $B^{(10)}$ of eq.
(\ref{10b2a}) (having an index in the 9-direction) is consistently
projected out\footnote{In the case of $B^{\prime(10)}$ defined in
eq. (\ref{10b'2a}), consistency would require that this form survive
the projection.}. In order to make the analogy with the IIB case in
10 dimensions manifest, we define the 10-dimensional $\G$-matrices
as
\bea & & \G^\m = \g^\m \otimes \sigma_2 \quad , \nonumber \\
& & \G^9 = {\bf 1} \otimes \sigma_1 \quad , \nonumber \\ & & \G_{11}
= {\bf 1} \otimes \sigma_3 \eea in terms of the 9-dimensional
$\g$-matrices. Consequently, denoting the 10-dimensional IIA spinors
as doublets of 9-dimensional spinors, the truncation acts as \bea &
& (1 \pm \sigma_1 ) \psi_\m =0 \quad , \nonumber \\ & & ( 1 \mp
\sigma_1 ) \l =0 \quad . \eea

We now want to consider the introduction of D8-branes, and we will
only take into account the case in which a single D8-brane is
located at one of the two fixed points of the orientifold
projection. Consistency requires that the truncation acts on the
world-volume fields as\footnote{The other world-volume field, the
scalar $x^9$, is of course projected out because of eq.
(\ref{ori9}).}
\bea & & V^i =0 \quad , \nonumber \\
& & ( 1 \pm \sigma_1 ) \theta =0 \quad .\eea  The brane action
contains, in a massive background, an additional Chern-Simons term
\cite{massive1,massive2} that we will not take into account because
it vanishes after the truncation. The relevant terms in the brane
action are thus \be -\int_{M_9} e^{-\phi} \sqrt {- \det g } +
\int_{M_9} C^{(9)} \quad . \ee The supersymmetrization of this
action is obtained in the Monge gauge following the same arguments
of the previous section. Taking into account only the terms that are
relevant after the truncation, and neglecting higher order fermi
fields, we thus define the hatted fields \bea & & \hat{\phi} = \phi
+ \2 \bar{\theta} \l +... \quad , \nonumber \\ & & \hat{g}_{\m\n} =
g_{\m\n} + 2 \bar{\theta} \g_{(\m} \psi_{\n)} + \bar{\theta}
\g_{(\m} D_{\n)} \theta +... \quad , \nonumber
\\ & & \hat{C}^{(9)}_{\m_1 ... \m_9}= C^{(9)}_{\m_1 ...\m_9} - 9
e^{-\phi} \bar{\theta} \G_{[ \m_1 ... \m_8} \G_{11} \psi_{\m_9 ]} -
\2 e^{-\phi} \bar{\theta} \G_{\m_1 ... \m_9} \G_{11} \l \nonumber \\
& & \qquad \qquad - \frac{9}{2} e^{-\phi} \bar{\theta} \G_{[\m_1 ...
\m_8} \G_{11} D_{\m_9 ]} \theta +... \quad , \eea whose
supersymmetry transformation has the form of a $\theta$-dependent
general coordinate transformation (plus an additional gauge
transformation in the case of the 9-form). Expressing then the brane
action in terms of these hatted fields, it turns out that if one
chooses the upper sign in the projection of the fermions, all the
terms containing the goldstino $\theta$ disappear in the action,
while the lower sign choice leads to an action of the VA type for
$\theta$. We interpret this result in  the same way as we did for
the IIB case in 10 dimensions. The upper sign choice corresponds to
a supersymmetric spectrum. Again, just as in the case of a single
D9-brane, for a single D8-brane there are no remaining world-volume
degrees of freedom after the truncation, but the generalization to a
stuck of branes would result in a spectrum in which supersymmetry is
linearly realized, and the goldstino is projected out. The resulting
action contains a dilaton tadpole and a RR tadpole, that in
consistent supersymmetric  orientifold models are both canceled
against the orientifold plane contribution. The other choice,
instead, corresponds to the case in which the brane and the
orientifold plane have both positive tension. Consequently, ${\cal
N}=1$ supersymmetry is non-linearly realized on the brane, and a
suitable orientifold projection only cancels the brane RR charge,
while a dilaton tadpole remains. The fact that this result is in
agreement with the IIB result of refs. \cite{truncation,fr} is a
manifestation of T-duality \cite{massive1,tduality}.

\section{S-duality of Type-IIB}\label{sec:s-duality}
Type-IIB superstring theory is conjectured to be invariant under
$SL(2,Z)$ transformations \cite{sduality}, a discrete subgroup of
the isometry group $SL(2,R)$ of type-IIB supergravity
\cite{schwarz}. This group acts on the complex scalar \be \tau = C_0
+i e^{-\phi} \ee as \be \tau \rightarrow \frac{a \tau +b}{c \tau +d}
\quad , \ee where
\be \left(%
\begin{array}{cc}
  a & b \\
  c & d \\
\end{array}%
\right) \in SL(2,R) \quad ,\ee while the 2-forms $B^{(2)}$ and
$C^{(2)}$ transform as a doublet. The matrix \be S =
\left(%
\begin{array}{cc}
  0 & 1 \\
  -1 & 0 \\
\end{array}%
\right) \ee generates the S-duality transformation $\tau \rightarrow
-\frac{1}{\tau}$, that for a vanishing axion background corresponds
to $\phi \rightarrow -\phi $, and in type-IIB string theory this
results in mapping weak coupling to strong coupling. $SL(2,Z)$
symmetry thus implies a strong-weak coupling self-duality of
type-IIB string theory. Since $S$ maps $B^{(2)}$ to $C^{(2)}$ and
viceversa, the duality interchanges the fundamental string and the
NS5-brane with the D1-string and the D5-brane. We want to study here
how spacetime-filling branes transform under S-duality, and since
the 10-forms $B^{(10)}$ and $C^{(10)}$ cannot appear consistently in
the low-energy effective action, the only way of deducing their
behavior under an $S$-transformation is to make use of the
supersymmetry algebra. In the remaining of this section, we thus
study how a transformation $S$ acts on the supersymmetry algebra
(\ref{susytransf}).

In the string frame, S-duality acts on the metric as
\be g_{\m\n}
\rightarrow e^{-\phi} g_{\m\n} \quad .
\ee
Because of the explicit
dilaton dependence of this transformation, it is easier to consider
a configuration with vanishing axion background. This is what we
will do in the following, and it is understood that our results do
not depend on this assumption. For completeness, we write again the
IIB supersymmetry transformations in the string frame in this
background:
\bea
& & \d e_\m{}^a = \bar{\e} \G^a \psi_\m \quad , \nonumber \\
& & \d \psi_\m = D_\m \e -\frac{1}{8} H_{\m\n\r} \G^{\n\r} \sigma_3
\e + \frac{1}{48} e^\phi G_{\m_1 \m_2 \m_{3}}^{(3)} \G^{\m_1 \m_2
\m_{3}}
\G_\m \sigma_1 \e \quad , \nonumber \\
& & \d B_{\m\n}^{(2)} = 2 \bar{\e } \sigma_3 \G_{[\m} \psi_{\n ]} \quad , \nonumber \\
& & \d B^{(10)}_{\m_1 ...\m_{10}} = e^{-2 \phi} \bar{\e} \sigma_3 (
10
\G_{[ \m_1 ...\m_9 } \psi_{\m_{10}]} - \G_{\m_1 ... \m_{10}}  \l ) \quad , \nonumber \\
& & \d C^{(2)}_{\m \n} =- 2 e^{- \phi} \bar{\e} \sigma_1
\G_{ [ m }  (\psi_{\n ] }- \frac{1}{4} \G_{\n ]} \l )\nonumber \\
& & \d C^{(10)}_{\m_1 ...\m_{10}} =- 10 e^{- \phi} \bar{\e} \sigma_1
\G_{ [ \m_1 ... \m_{9}}  (\psi_{\m_{10}] }- \frac{1}{20} \G_{\m_{10}]} \l )\nonumber \\
& & \d \l = \de_\m \phi \G^\m \e -\frac{1}{12} H_{\m\n\r} \sigma_3
\G^{\m\n\r} \e -\frac{1}{12} e^\phi G^{(3)}_{\m \n\r}
\sigma_1 \G^{\m \n\r} \e \quad , \nonumber \\
& & \d \phi = \frac{1}{2} \bar{\e} \l  \quad ,\label{susytransf2}
\eea again neglecting higher order fermi terms in the
transformations of the fermions.

Our strategy will be to derive the transformations of the fields
under S-duality requiring that the supersymmetry algebra is
preserved. We already know that \bea & & \phi \rightarrow - \phi
\quad , \nonumber
\\ & & e_\m{}^a \rightarrow e^{-\phi /2} e_\m{}^a \quad . \label{sdualphie}
\eea We now obtain the transformations of $\psi_\m$, $\l$ and $e$
requiring that they are consistent with eqs. (\ref{sdualphie}), {\it
i.e.} imposing that the supersymmetry variation of the S-transformed
fields is still eq. (\ref{susytransf2}), up to other local symmetry
transformations of the type-IIB theory. We know from the
supersymmetry transformation of $\psi_\m$ that $\e$ and $\psi_\m$
must acquire the same dilaton dependence. Moreover, we expect that
all the fermions undergo an overall $SL(2,R)$ rotation determined by
a $2\times2$ unitary matrix $\Omega$. Finally, the transformation of
the gravitino can contain a term proportional to $\G_\m \l$. Hence,
imposing that the transformed vielbein has the correct supersymmetry
variation, one gets \bea & & \e \rightarrow e^{-\phi /4} \Omega \e
\quad \nonumber \\ & & \psi_\m \rightarrow e^{-\phi /4} \Omega
\psi_\m - \frac{1}{4} e^{-\phi /4} \Omega \Gamma_\m \l\quad ,
\label{sdualpsi} \quad . \eea It turns out that the supersymmetry
transformation of the vielbein is mapped to itself plus an
additional local Lorentz transformation of parameter \be \L^{ab} =
\frac{1}{4} e^{- \phi /2} ( \bar{\e} \G^{ab} \l ) \quad . \ee We
neglect this term when we study the S-transformations of the
supersymmetry variations of the fermions, since they would lead to
cubic fermi terms. The S-duality transformation of $\l$ is
straightforwardly obtained imposing that the transformed dilaton
varies according to (\ref{susytransf2}) under supersymmetry, and the
result is \be \l \rightarrow - e^{\phi/4} \Omega \l \quad .
\label{sduallambda} \ee

Proceeding this way, one realizes that the two 2-forms $B^{(2)}$ and
$C^{(2)}$ form an $SL(2,R)$ doublet, transforming as \bea & &
B^{(2)} \rightarrow  C^{(2)} \quad , \nonumber \\
& & C^{(2)} \rightarrow - B^{(2)}  \label{sdual2forms} \eea if \bea
& & \Omega^{-1} \sigma_3 \Omega = - \sigma_1 \quad , \nonumber
\\ & & \Omega^{-1} \sigma_1 \Omega = \sigma_3 \quad , \label{omega}
\eea whose solution is\footnote{The inverse choice for $\Omega$
corresponds to a sign change in the transformations of $C^{(2)}$ and
$B^{(2)}$.} \be \Omega =
e^{-i\pi \sigma_2 / 4} =\left(%
\begin{array}{cc}
  1/\sqrt{2} & -1/\sqrt{2} \\
  1/\sqrt{2} & 1/\sqrt{2} \\
\end{array}%
\right) \quad . \ee Implementing these transformations on the
supersymmetry variation of $C^{(4)}$, one then obtains \be
C^{(4)}_{\m_1 ... \m_4} \rightarrow C^{(4)}_{\m_1 ... \m_4} - 6
B^{(2)}_{[ \m_1 \m_2} C^{(2)}_{\m_3 \m_4 ]} \quad , \ee leaving
invariant the field strength \be G^{(5)} = d C^{(4)} - H^{(3)}
\wedge C^{(2)} \quad . \ee The correctness of the transformations
(\ref{sdualpsi}) and (\ref{sduallambda}) is finally proven by
showing that the supersymmetry variation of the transformed fermi
fields is consistent with eqs. (\ref{susytransf2}).

Following the same arguments, one can now determine the S-duality
transformations of the two 10-forms $B^{(10)}$ and $C^{(10)}$. One
expects these fields to transform as a doublet, but the surprising
result is that the first requirement one has to make to impose
S-duality is that the transformation of this doublet must have a
non-trivial dilaton dependence. More precisely, the only possibly
consistent transformation is \bea & &
B^{(10)} \rightarrow e^{-2 \phi} C^{(10)} \quad , \nonumber\\
& & C^{(10)} \rightarrow - e^{-2 \phi} B^{(10)} \quad .
\label{sdual10forms} \eea This still does not guarantee that
S-duality is preserved, and in fact the additional variation of
$\phi$ in (\ref{sdual10forms}) is canceled only once one imposes the
additional constraints\footnote{As a consistency check, one can show
that these constraints are related by an S-duality transformation.}
\bea & & C^{(10)}_{\m_1 ... \m_{10}} ( \bar{\e} \l ) = - e^{-\phi}
(\bar{\e } \sigma_1 \G_{\m_1
...\m_{10}} \l ) \quad , \nonumber \\
& & B^{(10)}_{\m_1 ... \m_{10}} ( \bar{\e} \l ) = e^{- 2 \phi}
(\bar{\e } \sigma_3 \G_{\m_1 ...\m_{10}} \l ) \quad .
\label{constraint} \eea After performing the type-I truncation of
eq. (\ref{fermitrunc}), in which $B^{(10)}$ is projected out, the
first constraint becomes \be \frac{1}{10!} \e^{\m_1 ...\m_{10}}
C_{\m_1 ...\m_{10}} = \mp e^{-\phi} \sqrt{- \det g}\quad . \ee A
similar result holds for the second constraint, after performing the
heterotic truncation, in which all the RR fields are projected out
and the spinors are projected according to \be (1 \pm \s_3 ) f =0
\quad . \ee  As we will see in the next section, a possible
interpretation of this result is that S-duality is actually broken
in the presence of spacetime-filling branes. The constraint of eq.
(\ref{constraint}) can be justified only in the truncated theory,
and this is in agreement with the fact that only in the truncated
theory the presence of spacetime-filling branes can be consistent.

We will see in the next section that if one tries to construct the
S-dual of a supersymmetric D9-brane using eqs. (\ref{sdualphie}),
(\ref{sdualpsi}), (\ref{sduallambda}), (\ref{sdual2forms}) and
(\ref{sdual10forms}), the action one gets is no longer
$\kappa$-symmetric, and the constraints of eq. (\ref{constraint})
have to be imposed to restore $\kappa$-symmetry. This means that the
breakdown of $\kappa$-symmetry is consistent with the breakdown of
S-duality. The brane action obtained in this way has a DBI term
proportional to $e^{-4 \phi}$ and a WZ term proportional to $e^{-2
\phi}$. As we are going to prove, it turns out that, in the
untruncated theory, a $\kappa$-symmetric action for a
spacetime-filling brane charged with respect to $B^{(10)}$ has an
$e^{-2 \phi}$ dilaton dependence in the DBI term, and no dilaton
factor in the WZ term.

\section{Spacetime-filling branes and S-duality}\label{sec:ns9}
In this section we want to describe the $\kappa$-symmetric
spacetime-filling branes that are charged with respect to the NS
10-forms of IIB and IIA supergravities. We start from the type-IIB
case, performing an S-duality transformation on the D9-brane action.
In the following of this section, we will always have in mind to
perform a $Z_2$ truncation that leaves the NS 10-form invariant. In
the case of IIB, this transformation projects out all the RR fields,
leaving the NS fields invariant. This projection can actually be
worked out using the results of the previous section, performing an
S-duality transformation of the truncations of eq.
(\ref{fermitrunc}). In the fermionic sector, performing the
transformations (\ref{sdualpsi}) and (\ref{sduallambda}) and using
eq. (\ref{omega}), the projection becomes \be (1 \pm \s_3 ) f =0
\quad . \label{hettruncfer} \ee The same projection applies to the
spinor $\theta$ in the brane sector, since $\theta$ transforms under
S-duality like $\e$, while the world-volume vector $V^i$ is
projected out, again in agreement with eqs. (\ref{sdualpsi}) and
(\ref{omega}). We will assume that the 10-forms $B^{(10)}$ and
$C^{(10)}$ transform according to eq. (\ref{sdual10forms}), keeping
in mind that this is consistent only if the constraints
(\ref{constraint}) are satisfied. We will see that the action we end
up with is consistently $\kappa$-symmetric only if these constraints
are satisfied.

After the projection, the S-dual of the action (\ref{braneaction})
becomes  \be S =  - \int_{M_{10}} d^{10} \xi e^{- 4 \phi} \sqrt{ -
\det g } + \int_{M_{10}} e^{-2 \phi}  B^{(10)} \quad .
\label{nsbraneaction} \ee The supersymmetrization of this action is
then worked out in the same way as the D9 and D8 cases. One first
constructs the hatted fields \bea & & \hat{\phi} = \phi + \2
\bar{\theta} \l +... \quad , \nonumber \\ & & \hat{g}_{\m\n} =
g_{\m\n} + 2 \bar{\theta} \g_{(\m} \psi_{\n)} + \bar{\theta}
\g_{(\m} D_{\n)} \theta +... \quad , \nonumber \\ & &
\hat{B}^{(10)}_{\m_1 ... \m_{10}}= B^{(10)}_{\m_1 ...\m_{10}} + 10
e^{-2\phi} \bar{\theta} \G_{[ \m_1 ... \m_9} \s_3 \psi_{\m_{10} ]} -
e^{- 2\phi} \bar{\theta} \G_{\m_1 ... \m_{10}} \s_3 \l \nonumber \\
& & \qquad \qquad  +5 e^{-2\phi} \bar{\theta} \G_{[\m_1 ... \m_9}
\s_3 D_{\m_{10} ]} \theta +... \quad ,
\label{hattedfieldshettrunc}\eea and then writes the supersymmetric
action \be S =  - \int_{M_{10}} d^{10} \xi e^{- 4 \hat\phi} \sqrt{ -
\det \hat{g} } + \int_{M_{10}} e^{-2 \hat\phi} \hat{B}^{(10)} \quad
. \ee If this procedure preserved $\kappa$-symmetry, it would be
expected that one of the two truncations (the one with the upper
sign choice in eq. (\ref{hettruncfer}), as one would get using eq.
(\ref{omega})) leaded to a brane action in which all the goldstino
terms disappear. This is actually not the case, since for the upper
sign choice a term proportional to \be B^{(10)}_{\m_1 ... \m_{10}} (
\bar{\e} \l ) + e^{- 2 \phi} (\bar{\e } \G_{\m_1 ...\m_{10}} \l )
\ee survives. This term vanishes if the second constraint of eq.
(\ref{constraint}) is imposed. This is not surprising, since only if
this constraint is valid the S-duality transformations can be
performed. Thus, the picture that emerges is that the breakdown of
S-duality is in agreement with the breakdown of $\kappa$-symmetry,
and the constraint of eq. (\ref{constraint}) provides a restoration
of both. On the other hand, the constraint (\ref{constraint}) leads
to a vanishing action for the NS9-brane, and this could simply mean
that such an object does not exist. We will comment about this in
the conclusions. The lower sign choice in (\ref{hettruncfer}), again
analogously to the D-brane case, corresponds to a VA-type action for
$\theta$, after eq. (\ref{constraint}) is imposed.

Let us consider now the action \be S =  - \int_{M_{10}} d^{10} \xi
e^{- 2 \hat\phi} \sqrt{ - \det \hat{g} } + \int_{M_{10}}
\hat{B}^{(10)} \quad . \label{dualbraneaction} \ee In this case,
using eqs. (\ref{hattedfieldshettrunc}), one obtains that the upper
sign choice in eq. (\ref{hettruncfer}) leads to an action with no
goldstino, while the lower sign choice leads to a VA action, and in
both cases no constraint is required. This means that the
untruncated action is $\kappa$-symmetric, and thus we argue that
this is the correct action for an NS9-brane. More precisely, the
complete action would result from rescaling the S-dual of the
lagrangian of eq. (\ref{braneaction}), and in order to compute this,
one should know how $C^{(8)}$ and $C^{(6)}$ transform under
S-duality. This analysis is currently under investigation, and we
expect that it would shed some light on the problem of studying the
S-dual of a D7-brane as well. Anyway, we do not expect the results
of this section to be altered by the inclusion of additional terms,
since we do not see how the constraint of eq. (\ref{constraint}) can
be removed modifying $B^{(10)}$ by the inclusion of other bulk
fields.

One could also discuss the S-dual of this picture,  starting from
the action of eq. (\ref{dualbraneaction}), and then performing an
S-duality transformation. The result is that one ends up with the
action \be S = - \int_{M_{10}} d^{10} \xi e^{- 3 \hat\phi} \sqrt{ -
\det \hat{g} } + \int_{M_{10}} e^{-2 \hat\phi} \hat{C}^{(10)} \quad
. \label{braneactionphi3} \ee Again, $\kappa$-symmetry corresponds
to the existence of a $Z_2$ truncation removing the goldstino
completely, and one can show that this happens only after imposing
the constraint for $C^{(10)}$ in eq. (\ref{constraint}). The picture
is thus completely symmetric, since from this low-energy point of
view assuming that the D9-brane action is (\ref{braneaction})
instead of (\ref{braneactionphi3}) is S-dual to assuming that the
action for an NS9-brane is (\ref{dualbraneaction}) instead of
(\ref{nsbraneaction}).

At the end of this section, we want to determine the supersymmetric
action for an NS9-brane in type-IIA, where again $\kappa$-symmetry
corresponds to the vanishing of all the goldstino terms in the
suitably $Z_2$-truncated action. The 10-dimensional $Z_2$-truncation
projects out all the RR fields, acting on the fermions as \bea & &
\psi_\m = \pm \G_{11} \psi_\m \quad , \nonumber
\\ & & \l = \mp \G_{11} \l \quad . \label{fertrunc2a} \eea
From the supersymmetry transformation of eq. (\ref{10b2a}) we obtain
\bea & &  \hat{B}^{(10)}_{\m_1 ... \m_{10}} = B^{(10)}_{\m_1 ...
\m_{10}} -10 e^{-2 \phi} \bar{\theta} \G_{[ \m_1 ...\m_9 }
\psi_{\m_{10}]} + e^{-2 \phi} \bar{\theta} \G_{\m_1 ... \m_{10}} \l
\nonumber \\ & & \qquad \qquad -5 e^{-2 \phi } \bar{\theta}
\G_{[\m_1 ... \m_9 } D_{\m_{10}]} \theta + ... \quad . \eea The
truncation acts on the world-volume fields as usual: the vector
$V^i$ is projected out, while $\theta$ transforms in  the same way
as $\psi$. The final result is that the truncated action \be S =  -
\int_{M_{10}} d^{10} \xi e^{- 2 \hat\phi} \sqrt{ - \det \hat{g} } +
\int_{M_{10}} \hat{B}^{(10)} \ee is supersymmetric choosing the
upper sign in eq. (\ref{fertrunc2a}), while supersymmetry is
spontaneously broken if one chooses the lower sign. It can be shown
that, after an $S^1$ reduction, T-duality relates this action with
the one of eq. (\ref{dualbraneaction}).

\section{Conclusions}\label{sec:conclusions}
The starting point of this paper was a continuation of \cite{fr},
where the results of \cite{truncation} were generalized to a curved
background, showing that the possible type-I truncations of type-IIB
are in correspondence with the possible consistent type-I strings in
D=10. We showed here that the same results apply to the D=9
truncations of type-IIA, in accordance with T-duality. We then
proceeded constructing the $\kappa$-symmetric spacetime-filling
branes that are charged with respect to the NS 10-forms of type-IIB
and type-IIA.

In \cite{hull} it was argued that S-duality of type-IIB implies the
existence of NS9-branes, that together with the D9-branes form an
$SL(2,Z)$ doublet. From the standard Weyl-rescaling argument, it
turns out that the tension of these branes appears to scale like
$1/g_S^4$ in the string frame. Here we have argued that the actual
tension of these branes scales like $1/g_S^2$, like the other
solitonic NS objects, namely NS5-branes. The solution of the paradox
is that the doublet of NS and RR 10-form potentials does not
transform covariantly under $SL(2,Z)$. It is therefore not possible
to derive the action for an NS9-brane performing a Weyl rescaling.
In \cite{hull,hull2} it was also conjectured that S-duality implies
the existence of a dual of the orientifold projection, and the
$SO(32)$ heterotic theory should result from this projection, after
the introduction of 32 NS9-branes. This projection would naturally
act like $\sigma_3$ on the fermion doublets, since in the heterotic
theory the fermions come only from the left sector. We do not expect
this truncation to be an ordinary $Z_2$ orbifold, since it is well
know that a $Z_2$ orbifold of type-IIB gives rise to type-IIA. In
any case, if there is a way of deriving the heterotic $SO(32)$
theory from type-IIB, we expect that the `twisted' sector of the
projection would correspond to inserting $\kappa$-symmetric branes,
that would therefore have the structure of eq.
(\ref{dualbraneaction}). In any case, should a brane interpretation
of the heterotic theory be possible, a natural question would arise,
namely what is the heterotic string equivalent of brane
supersymmetry breaking\footnote{I am grateful to E. Dudas for
discussions about this point.}.

Similar arguments hold for the IIA case. Since the type-IIA
superstring and the $E_8 \times E_8$ heterotic theory have both an
M-theory origin \cite{witten,hw}, it has also been conjectured
\cite{hull} that the $E_8 \times E_8$ heterotic theory can arise in
ten dimensions from a projection of type-IIA, that would result in
the low-energy action in a $Z_2$ truncation removing the RR fields.
We emphasize again that if this projection exists, it cannot act as
a $Z_2$ orbifold of type-IIA, since such an orbifold gives rise to
type-IIB. The `twisted' sector of the heterotic theory would result
in this case form the insertion of NS9-branes. Starting from
type-IIB and using Weyl-rescaling arguments, it has been argued that
T-duality would imply that the tension of this branes, apart from
having an $e^{-4 \phi}$ dilaton dependance, is proportional to
$R^3$, where $R$ is the radius of the isometry direction
\cite{trunc2}, and this would mean that they are not defined in 10
uncompactified dimensions. The NS9-brane, as well as the D8-brane,
would then result from a 9-brane in M-theory whose effective action
and target space solution \cite{m9} can be written only if the
11-dimensional supergravity has an isometry, and thus cannot be
covariant in 11-dimensions. Again, the $\kappa$-symmetric
spacetime-filling brane we obtained in this paper has instead an
$e^{-2\phi}$ dilaton dependance, and it is related by T-duality to
the type-IIB NS-brane of eq. (\ref{dualbraneaction}). This different
scaling with respect to the one of \cite{trunc2} implies that this
NS9-brane and the D8-brane can not have a common M-theory origin.
Since the field-strength of a 10-form in 11 dimensions would be dual
to a cosmological constant, this result is basically rephrasing the
fact that no cosmological constant can be included in 11-dimensional
supergravity \cite{bdhs}, and Romans IIA supergravity can not be
obtained by dimensional reduction from 11 dimensions. After
compactification on a 2-torus M-theory is related to type-IIB by
T-duality, and thus this picture is the T-dual analogous of the
type-IIB picture, where S-duality is broken by the presence of
spacetime-filling branes.

Finally, it would be interesting to see if the S-duality rules of
this paper can be used to understand the strong coupling behavior of
the D7-branes. In \cite{mo} it was shown that D7-branes of type-IIB
belong to a triplet of 7-branes. One could then determine a
supersymmetric effective action for these branes requiring
$\kappa$-symmetry, and relate them to the half-BPS 7-brane solutions
of type-IIB supergravity \cite{ggp,epz,groningen}. This analysis is
currently under investigation.

\vskip 2cm

\section*{Acknowledgments}\label{sec:acknowledgments}
I am grateful to E. Bergshoeff, M. Bianchi, E. Dudas, M. Green and
G. Pradisi for discussions. This work is supported by a European
Commission Marie Curie Postdoctoral Fellowship, Contract
MEIF-CT-2003-500308.

\end{document}